\documentclass[onecolumn,12pt]{revtex4}
\def\gsim{ \lower .75ex \hbox{$\sim$} \llap{\raise .27ex \hbox{$>$}} }
\def\lsim{ \lower .75ex \hbox{$\sim$} \llap{\raise .27ex \hbox{$<$}} }
\usepackage{graphicx,latexsym,amsmath,amssymb}
\usepackage{bm} \input epsf \def\gsim{ \lower .75ex \hbox{$\sim$}
\llap{\raise .27ex \hbox{$>$}} } \def\lsim{ \lower .75ex \hbox{$\sim$}
\llap{\raise .27ex \hbox{$<$}} }

\newcommand{\be}{\begin{equation}}
\newcommand{\ee}{\end{equation}}
\newcommand{\bea}{\begin{eqnarray}}
\newcommand{\eea}{\end{eqnarray}}

\begin{document}
\title{The Return of the Phoenix Universe}
\author{Jean-Luc Lehners}
\email{jlehners@princeton.edu} \affiliation{Princeton Center
for Theoretical Science, Princeton University, Princeton, NJ
08544 USA}
\author{Paul J. Steinhardt}
\email{steinh@princeton.edu}
\affiliation{Princeton Center for
Theoretical Science and Joseph Henry Laboratories, Princeton
University,
Princeton, NJ 08544, USA}
\author{Neil Turok}
\email{nturok@perimeterinstitute.ca} \affiliation{Perimeter
Institute for Theoretical Physics, Waterloo,
Ontario,~Canada~N2L~2Y5
\\}

\date{30th March 2009}

\begin{abstract}
Georges Lemaitre introduced the term {\it phoenix universe} to
describe an oscillatory cosmology with alternating periods of
gravitational collapse and expansion. This model is ruled out
observationally because it requires a supercritical mass
density and cannot accommodate dark energy. However, a new
cyclic theory of the universe has been proposed that evades
these problems. In a recent elaboration of this picture, almost
the entire universe observed today is fated to become entrapped
inside black holes, but a tiny region will emerge from these
ashes like a phoenix to form an even larger smooth, flat
universe filled with galaxies, stars, planets, and, presumably,
life. Survival depends crucially on dark energy and suggests a
reason why its density is small and positive today.
\\ \\ {\it Essay written for the Gravity Research Foundation 2009
Awards
for Essays on Gravitation.}
\end{abstract}

\maketitle

\vspace{5cm}

\begin{verse} {\it ``those solutions where the universe expands and
contracts successively [...], have an indisputable poetic charm
and make one think of the phoenix of legend.''  }
\\ Georges Lemaitre, 1933 \cite{Lemaitre}
\end{verse}

Two breakthroughs of the twentieth century changed forever our
understanding of the universe: the observation that the
universe is expanding, made in the 1920s, and that the
expansion rate is accelerating, made in the 1990s. The full
implications have yet to be realized. The currently favored
inflationary picture does not explain the origin of the
expansion -- the big bang -- or provide a rationale for the
current acceleration. Recently, though, a new cosmological
model has emerged that breathes new life into an old idea --
the {\it phoenix universe} -- providing an explanation for both
the bang and the dark energy, and suggesting why the latter
must be small and positive today.

The ``phoenix'' was first introduced into cosmology by Georges
Lemaitre shortly after Hubble's discovery that the universe is
expanding.  Friedmann and Lemaitre had discussed the expanding
universe model several years earlier, but its realization in
nature forced cosmologists to face up to its baffling
beginning: the big bang, the moment about fourteen billion
years ago when the temperature and density reached infinite
values.  The standard interpretation today is that the bang
marked the beginning of space and time. However, this is far
from proven: all we really know is that Einstein's equations
fail and an improved theory of gravity is needed. In fact, the
idea of a ``beginning,'' the emergence of the universe from
nothing, is a very radical notion. A more conservative idea is
that the universe existed before the big bang, perhaps even
eternally. Historically, this motivated many of the founders of
the big bang theory, including Friedmann, Lemaitre, Einstein
and Gamow, to take seriously an ``oscillatory'' universe model
in which every epoch of expansion is followed by one of
contraction and then by a ``bounce,'' at an event like the big
bang, to expansion once more. For the model to work, the matter
must exceed the critical density required for its
self-attraction to slow the expansion and eventually reverse it
to contraction. But by the end of the twentieth century,
observations had shown the opposite: the matter density is
subcritical and the expansion is speeding
up~\cite{Komatsu:2008hk}.

Yet, today, the phoenix universe has been revived due to the
development of a new cyclic theory of the universe that
incorporates dark energy and cosmic acceleration in an
essential way \cite{Steinhardt:2002ih}. To explain the theory,
it is useful to invoke a picturesque version inspired by string
theory and M-theory in which space-time consists of two
three-dimensional braneworlds separated by a tiny gap along an
additional spatial dimension. One of these braneworlds is the
world we inhabit. Everything we can touch and see is confined
to our braneworld; the other is invisible to us. According  to
this picture, the big bang corresponds to a collision between
the braneworlds, followed by a rebound. Matter, space and time
exist before as well as after, and it is the events that occur
before each bang that determine the evolution in the subsequent
period of expansion.

Unlike Lemaitre's phoenix universe, the matter density is
subcritical, consistent with observations.  The big bang
repeats at regular intervals because a spring-like force keeps
drawing the braneworlds together along the extra dimension,
causing them to collide every trillion years or so.  Associated
with the spring-like force are kinetic and potential energy,
which play an important role as the source of dark energy in
the cyclic model.

The dark energy equation of state $w$ is defined as the ratio
of the pressure (kinetic minus potential energy of the
braneworlds) to the total energy density (kinetic plus
potential energy).  When the braneworlds are farthest apart,
the total energy is predominantly potential and positive,
corresponding to $w \approx -1$, similar to a cosmological
constant.  Although this potential energy is negligible right
after a collision, it decreases slowly and, about nine billion
years later, overtakes the matter density, causing the
expansion of the braneworld to accelerate. The acceleration
cannot last forever, though, because the spring eventually
releases, causing the braneworlds to hurtle towards each other.
Now, the potential energy decreases and becomes negative while
the kinetic energy grows, causing $w$ to increase sharply from
$w \approx -1$ to $w \gg 1$ and initiating a period known as
``ekpyrosis''~\cite{Khoury:2001wf}. From the point of view of a
``braneless observer,'' someone who is unaware of the extra
dimension and the other braneworld and reinterprets the
goings-on in terms of usual Einstein general relativity, the
universe appears to be undergoing a peculiar period of
ultra-slow contraction in which the scale factor $a(t) \sim
(t_{bang}-t)^{2/3(1+w)}$ as $t$ approaches $t_{bang}$ with
$w\gg1$. The dark energy continues to dominate the universe
during this ekpyrotic contraction phase, and the matter density
remains negligibly small.

The ekpyrotic phase is key, because it removes any need for
inflation. The horizon problem is resolved simply because the
universe exists long before the big bang, allowing distant
regions to become causally connected.  To see how the flatness
puzzle is solved without inflation, recall that the problem
arises in a slowly expanding universe, where a small deviation
from flatness at early times grows into an unacceptably large
one by the present epoch.  But now just run the story
backwards: as space slowly contracts, an initially large
deviation from flatness shrinks to an infinitesimal one.  In an
ekpyrotic contraction phase, because $w\gg1$, the deviation
from flatness is diminished by more than it grows during the
subsequent expansion phase, thus explaining why it is
negligibly small today \cite{warn}.

Both ekpyrotic contraction and inflation can generate large
scale density fluctuations from microscopic quantum
fluctuations.  In inflation this occurs because quantum
fluctuations are stretched exponentially while the Hubble
horizon increases very slowly, so the fluctuations end up
spanning superhorizon scales.  In the ekpyrotic contraction
phase, the same feat is accomplished because the quantum
fluctuations remain nearly fixed in scale while the Hubble
horizon shrinks rapidly. By the time the phase ends, quantum
fluctuations formed inside the horizon span superhorizon
scales, resulting in a  spectrum of nearly scale-invariant
fluctuations very similar to inflation, although with
observably different predictions for primordial gravitational
waves \cite{Boyle:2003km} and non-Gaussian density fluctuations
\cite{Lehners:2007wc}.

An important caveat arises, though, for the best understood
example of ekpyrosis, where the density perturbations are
generated by a so-called entropic mechanism
\cite{Lehners:2007ac}. The ekpyrotic energy only maintains $w
\gg 1$ if the quantum fluctuations remain within a narrow
range. Otherwise, $w$ drops precipitously, inhomogeneities and
curvature grow, and space collapses into a warped amalgamation
of black holes. The chance of avoiding decimation is small:
during every e-fold of contraction, quantum fluctuations reduce
the fraction of space with $w \gg 1$ by $1/e$.  Since the
ekpyrotic phase lasts for about 120 e-folds, the fractional
volume of space that makes it smoothly to the bounce and
re-emerges in a flat, expanding phase is  $f \approx e^{-360}$
\cite{Lehners:2008qe}.  This fraction is so tiny that, if the
ekpyrotic phase started today, fourteen billion years after the
big bang, the entire observable universe ($10^{84}$~cm$^3$
across) would be decimated.

Dark energy saves the universe from this ashen fate by causing
the expansion to accelerate. If acceleration continues for at
least 560 billion years ($> 56$ e-folds), a volume of at least
a cubic centimeter will retain its $w \gg 1$ ekpyrotic form all
the way to the next crunch and emerge unscathed: flat, smooth
and isotropic.  As tiny as a cubic centimeter may seem, it is
enough to produce a flat, smooth region a cycle from now
 at least as large as the region we currently observe.
In this way, dark energy, the big crunch and the big bang all
work together so that the phoenix forever arises from the
ashes, crunch after crunch after crunch.

The revival of the phoenix universe could also resuscitate an
old proposal for solving one of the deepest mysteries in
science: why the cosmological constant (or, equivalently, the
dark energy density when $w \approx -1$) is $10^{120}$ times
smaller than dimensional analysis suggests. The proposal
involved introducing a mechanism which causes the cosmological
constant to relax to smaller values. Starting out large, it
naturally decreases but its downward drift slows dramatically
as it becomes small. Should it ever slip below zero,
gravitational collapse follows swiftly. The result is that, for
a vast majority of the time and throughout almost all of space,
the cosmological constant is tiny and positive, just as we
observe.

Attempts to incorporate this idea into models where the big
bang is the beginning failed because the relaxation process
takes vastly longer than fourteen billion years.  There is
plenty of time in a cyclic universe, though. The relaxation can
occur without disrupting the cycles and vice-versa, so that an
overwhelming majority of cycles occur when the cosmological
constant is small and positive \cite{Steinhardt:2006bf}. By
incorporating the effects of dark matter, ordinary matter and
radiation on the rate of drift, it may even be possible to
explain the quantitative value observed today.


\begin{thebibliography}{99}
\bibitem{warn} To complete the argument, note that, as shown in
    J.~K.~Erickson, D.~H.~Wesley, P.~J.~Steinhardt and N.~Turok,
  Phys.\ Rev.\  D {\bf 69}, 063514 (2004), not only curvature, but also anisotropies are
suppressed in a contracting phase with  $w > 1$.
\end{thebibliography}
\end{document}